\documentclass[amsmath,amsfonts,aps,prd]{revtex4}
\usepackage[dvips]{color}
\usepackage[dvipdfm]{hyperref}
\usepackage{latexsym}
\usepackage{epsfig}
\usepackage{xcolor}

\newcommand {\mpl}{M_\mathrm{Pl}}
\newcommand{\be}{\begin{eqnarray}}
\newcommand{\ee}{\end{eqnarray}}

\newcommand{\bo}{\raise-1mm\hbox{\Large$\Box$}} 

\begin{document}

\title{Standard Cosmological Evolution in a Wide Range of $f(R)$ Models}
\author{Jonathan D. Evans}
\email[Email address: ]{masjde@maths.bath.ac.uk}
\affiliation{Department of Mathematical Sciences, University of Bath, Bath BA2 7AY, United Kingdom}
\author{Lisa M.H. Hall}
\email[Email address: ]{Lisa.Hall@sheffield.ac.uk}
\affiliation{Astroparticle Theory and Cosmology Group, Department of Applied Mathematics, The University of Sheffield, Hounsfield Road, Sheffield S3 7RH, United Kingdom}
\author{Philippe Caillol}
\email[Email address: ]{p.l.caillol@sheffield.ac.uk}
\affiliation{Solar Physics and Space Plasma Research Centre, Department of Applied Mathematics, The University of Sheffield, Hounsfield Road, Sheffield S3 7RH, United Kingdom}

\begin{abstract} 
Using techniques from singular perturbation theory, we explicitly calculate the cosmological evolution in a class of modified gravity models.
By considering the (m)CDTT model, which aims to explain the current acceleration of the universe with a modification of gravity, we show that Einstein evolution can be recovered for most of cosmic history in at least one $f(R)$ model.
We show that a standard epoch of matter domination can be obtained in the mCDTT model, providing a sufficiently long epoch to satisfy observations.
We note that the additional inverse term will not significantly alter standard evolution until today and that the solution lies well within present constraints from Big Bang Nucleosynthesis.
For the CDTT model, we analyse the ``recent radiation epoch'' behaviour ($a\propto t^{1/2}$) found by previous authors.
We finally generalise our findings to the class of inverse power-law models.
Even in this class of models, we expect a standard cosmological evolution, with a sufficient matter domination era, although the sign of the additional term is crucial.
\end{abstract}

\maketitle

\section{Introduction}

The observation that the universe appears to be accelerating at present times has caused one of the greatest problems to modern cosmology.  
High precision data from Type Ia supernovae, the Cosmic Microwave Background and large scale structure seem to hint that the universe is presently dominated by an unknown form of energy, dubbed {\it dark energy} \cite{Riess:1998cb,Perlmutter:1998np,Tonry:2003zg,Bennett:2003bz,Netterfield:2001yq,Halverson:2001yy}.
The problem resides in the lack of good physical motivation for this energy density.  
One obvious contender for the role of dark energy is Einstein's cosmological constant, but particle physics fails to predict the correct density.
Other candidates include scalar fields or extra dimensions.  We refer the reader to \cite{Copeland:2006wr} for a review.

Recently, a modification of General Relativity itself was suggested to explain this accelerating universe \cite{Deffayet:2001pu,Capozziello:2002rd,Freese:2002sq,Arkani-Hamed:2002fu,Dvali:2003rk,Nojiri:2003ni,Arkani-Hamed:2003uy,Carroll:2003wy,Capozziello:2003tk,
Nojiri:2003ft,Carroll:2004de,Abdalla:2004sw,Capozziello:2006dj,Appleby:2007vb}.
For reviews see e.g. \cite{Easson:2004fq,Nojiri:2006ri,Trodden:2006qk}.
The postulation is that in which the Ricci scalar of the Einstein-Hilbert lagrangian is replaced by an arbitrary function of the Ricci scalar, $f(R)$.
The most simple example includes inverse powers of $R$: $f(R)=R-a/R^n$~\cite{Carroll:2003wy}.  
While this modified action was shown to give an accelerated attractor, due to the equivalence to scalar-tensor theories with a vanishing Brans-Dicke parameter\cite{Chiba:2003ir,Chiba:2006jp,Erickcek:2006vf}, solar system experiments rule these models out ($\omega_{BD}>40,000$~\cite{Bertotti:2003rm}).  Attempts have been made to overcome this problem~\cite{Nojiri:2003ft}, but solar system gravity still raises doubt as to these models~\cite{Brookfield:2006mq}. 

These models have raised much recent interest, due to their perhaps simple nature.  The presence of ghosts and stabilities have been studied \cite{Dick:2003dw,DeFelice:2006pg,Faraoni:2005vk}, as well as the implications to the CMB, large scale structure, solar system gravity and gravitational waves
~\cite{Amarzguioui:2005zq,Cembranos:2005fi,Song:2006ej,Navarro:2006mw,Bean:2006up,
Faulkner:2006ub,Li:2007xn,Fairbairn:2007qp,
Hu:2007nk,DeFelice:2007zq,Starobinsky:2007hu,
DeFelice:2007ez,Song:2007da,
Capozziello:2007ms,Pogosian:2007sw,
Carloni:2007yv,Ananda:2007xh}. 

Recently, it has been claimed that, in all theories which behave as a power of $R$ at large or small $R$, standard cosmological evolution can not be obtained~\cite{Amendola:2006kh,Amendola:2006eh}.
Further, it is claimed that no $f(R)$ will lead to an acceptable matter era\cite{Amendola:2006we}.
Specifically, during the matter phase, the authors find that in these theories the scale factor evolves as $a\propto t^{1/2}$ and not as $t^{2/3}$ as required.
In order for a modified theory of gravity to be convincing, we require a close-to-Einstein evolution in the past, with a standard, sufficiently-long period of matter domination.  We note that several models have been proposed which lead to matter dominance before acceleration~\cite{Nojiri:2006gh,Nojiri:2006be,Hu:2007nk} and that this issue is a matter of debate in the literature.

If the modification to gravity exists as a perturbative term in addition to the Einstein-Hilbert terms, then it is the belief of the authors of this paper that Einstein cosmology should be recovered in the limit that the perturbation term disappears.
This system has great similarity with perturbation theory~\cite{Hinch}.
This paper studies the asymptotic expansion of the (m)CDTT model \cite{Carroll:2003wy}, to show that standard cosmological evolution can be produced in at least one model.  
Specifically, we aim to recover an acceptable matter era and show that there is at least a class of $f(R)$ for which this is possible. 
For the purposes of this paper, we do not concern ourselves with the constraints from solar system experiment.  

The paper is laid out as follows.  In Section \ref{sec:theory}, we define the theoretical framework and give some useful formulae.  We state the simplest example in Section \ref{sec:CDTT} and derive the asymptotic solution in Section \ref{sec:asymp}.  
The relevance of this solution to cosmology is highlighted in Section \ref{sec:relevance}, where we comment on the era of matter domination, as well as Big Bang Nucleosynthesis.  After generalising our specific example to the class of inverse power-law models in Section \ref{sec:gen}, we conclude in Section \ref{sec:conc}.  We discuss the eigenmode analysis in the Appendix.

\section{Theoretical Framework}
\label{sec:theory}
For convenience, we consider a class of modified gravity in which we add a perturbative function, $f(R)$, to the Einstein-Hilbert action.  
We consider the action
\begin{equation}
{\cal S}_{\rm JF} = \int d^4 x \sqrt{-g}\left(\frac{R+f(R)}{2\kappa^2} + {\cal L}_{\rm m}\right),
\label{eq:action}
\end{equation}
where $R$ is the Ricci scalar and ${\cal L_{\rm m}}$ is the matter Lagrangian. 
The field equations, using the metric approach, can be derived from this action,
\begin{equation}\label{fieldequation}
R_{\mu\nu}-\frac12 g_{\mu\nu}R+{f_{R}}R_{\mu\nu}-\left(\frac{f}{2}-\bo f_{R}\right)g_{\mu\nu} -\nabla_{\mu}\nabla_{\nu}f_{R} = \kappa^{2} T_{\mu\nu}
\end{equation}
where $R_{\mu\nu}$ and $T_{\mu\nu}$ are the Ricci and stress-energy tensors respectively and we define $f_{R} \equiv df/dR$.  Later, we will also require $f_{RR} \equiv d^{2}f/dR^{2}$.
Although this theory can be rewritten as a scalar tensor theory (e.g. \cite{Flanagan:2003iw,Nojiri:2003ft}), we shall not use the conformal transformation; we complete all our calculations in the Jordan frame, given by the action above.

In a flat Friedmann-Robertson-Walker (FRW) background, the Ricci scalar can be written in terms of the scale factor, $a(t)$ and the Hubble parameter, $H=\dot a / a$.  
An overdot denotes differentiation with respect to cosmic time.
Specifically, 
\begin{eqnarray}
R=6\left(\frac{\ddot a}{a}+\frac{\dot a^2}{a^2}\right)=6\left(\dot H+2H^{2}\right).
\end{eqnarray}
In this modified theory, the Friedmann equation becomes
\begin{eqnarray}
3H^{2}-3f_{R}\left(\dot H +H^{2}\right) +\frac{f}{2} +18 H f_{RR}\left(\ddot H +4H\dot H\right) = \kappa^{2}\rho.
\label{eq:fried}
\end{eqnarray}
As expected, when $f(R)=0$, we recover Einstein cosmology, $3H^2=\kappa^2\rho$.  The continuity equation remains the same as in Einstein gravity:
\begin{eqnarray}
\dot \rho_T +3H \left(\rho_T+p_T\right) = 0,
\end{eqnarray}
where $\rho_T$ and $p_T$ are the total energy density and pressure respectively.  These can be separated into components of radiation and matter, for example $\rho_T=\rho_r+\rho_m$. 

We assume the parameter change as in CDDETT~\cite{Carroll:2004de}:
\begin{eqnarray}
x = -H(t) \qquad y = {\dot H}(t) \nonumber 
\end{eqnarray}
so that
\begin{eqnarray}
\ddot H = -y \frac{dy}{dx} = -\frac{2x^3}{v^2} + \frac{x^4}{v^3}\frac{dv}{dx}, \qquad \qquad R = \frac{6x^2(2v-1)}{v}.
\end{eqnarray}
As in CDDETT, we also define a function, $v(x)$,
which can be related to the power of a power-law solution, $a(t)\propto t^p$:
\begin{eqnarray}
v(x) = -\frac{x^2}{y} \nonumber
\end{eqnarray}
Then, power-law solutions can be identified simply, since $v(x)\rightarrow p ={\rm constant}$.  
In standard Einstein cosmology, solutions for $v(x)$ are expected to plateau at $v=\frac12$ and $v=\frac23$, which correspond to epochs of radiation and matter domination respectively.  We also note that an accelerating phase must have $v(x) >1$.
Using these new variables, it is then possible to rewrite Eqn. (\ref{eq:fried}) in terms of $x$,$v$ and $dv/dx$.
This equation is often simpler to solve, as found in CDDETT.

One most natural way of maintaining Einstein cosmology in the past whilst also explaining late-time acceleration is to consider a perturbative function, $f(R)$, (perhaps an inverse power-law) which incorporates some perturbation parameter.  It will be convenient later to separate this small parameter from the remaining function:
\begin{eqnarray}
f(R) = \epsilon \tilde f(R).
\label{eq:term_eps}
\end{eqnarray}
In terms of this new function, $\tilde f(R)$, (and using the new variables, $x$ and $v$) the Friedmann equation takes the form:
\begin{eqnarray}
3x^{2}-\kappa^{2}\rho_T = \epsilon s(x,v) \frac{dv}{dx} + \epsilon t(x,v)
\label{eq:fried_eps}
\end{eqnarray}
where
\begin{eqnarray}
t &=& \frac{3x^{2}(v-1)}{v}\tilde f_{R} -\frac{\tilde f}{2} +\frac{36 x^{4}(2v-1)}{v^{2}} \tilde f_{RR} \nonumber \\
s &=& \frac{18x^{5}}{v^{3}} \tilde f_{RR} \nonumber
\end{eqnarray}
Eqn. (\ref{eq:fried_eps}) can be rearranged to give the useful formula:
\begin{eqnarray}
\epsilon \frac{dv}{dx} = -\epsilon \frac{t(x,v)}{s(x,v)} + \frac{\left[3x^{2}-\kappa^{2}\rho_T\right]}{s(x,v)}.
\label{eq:generaldiff}
\end{eqnarray}

\section{Example: The CDTT Model}
\label{sec:CDTT}
One of the simplest modifications which aims to give late-time acceleration was proposed in \cite{Carroll:2003wy} (the CDTT model):  
\begin{eqnarray}
f(R) = \sigma \frac{\mu^{4}}{R}.
\end{eqnarray}
In the original paper, $\sigma=-1$.  We will also consider the mCDTT model, where $\sigma=1$.
We do not wish to advocate this model as a realistic modification of gravity, due to the impossibility to reconcile this action with solar system gravity~\cite{Chiba:2003ir}.  
Additionally, in order to justify the observation of a presently accelerating universe (the coincidence problem), we require $\mu\sim H_0$, which is fine-tuned.
Ignoring these problems, we use this simple model to demonstrate the techniques.

From Eqn.~(\ref{eq:fried}), the dynamical equations in such an action are
\begin{eqnarray}
\label{eq:newfriedmann}
3H^2 + \frac{\sigma\mu^4}{12({\dot H}+2H^2)^3}\left(2H{\ddot H}
+15H^2{\dot H}+2{\dot H}^2+6H^4\right) = \kappa^2\rho \nonumber.
\end{eqnarray}
We always consider the general case with both radiation and matter, such that $\rho=\rho_r+\rho_m$.
In the terminology of Eqn.~(\ref{eq:term_eps}), $\epsilon=\mu^4$ and $\tilde f(R) =\sigma 1/R$.  
We therefore obtain the system:
\begin{eqnarray}
\mu^4\frac{dv}{dx}&=&\mu^4 \frac{3(1-2v)(2-v)v}{2x}-6\sigma x(2v-1)^3\left[\kappa^2(\rho_r+\rho_m)-3x^2\right] \label{eq:fullv}\\
\frac{d\rho_r}{dx} &=& \frac{4v\rho_r}{x} \label{eq:fullrad}\\
\frac{d\rho_m}{dx} &=& \frac{3v\rho_m}{x} \label{eq:fullmat}.
\end{eqnarray}
Trivially, when $\mu=0$, we have:
\begin{eqnarray}
\kappa^2(\rho_r+\rho_m)-3x^2=0,\quad {\rm or} \quad (2v-1)=0 \nonumber
\end{eqnarray}
We ignore the solution, $v=1/2$, which is picked up due to rearrangement of the Friedmann equation, Eqn.~(\ref{eq:newfriedmann}).  With only radiation present ($\rho_m=0$) and using Eqn.(\ref{eq:fullrad}), $v=1/2$.  With only matter ($\rho_r=0$), $v=2/3$.
We, of course, wish to study the asymptotical analysis, when $\mu \ne 0$, which will be the focus of the next section.

\section{Asymptotic Analysis of the CDTT Model}
\label{sec:asymp}

We now study the CDTT model in the case when $\mu \ne 0$, but is a perturbative parameter.  We wish to find the asymptotic solution as both $\mu\to 0$ (to compare to Einstein's solution in the past) and $x=O(\mu)$ (with the aim to find late-time acceleration).
In the limit $\mu \to 0 $, (\ref{eq:fullv})--(\ref{eq:fullmat}) gives a singular perturbation problem, the solution to which we describe using the method of matched asymptotic expansions
\cite{Hinch}.
The solution on the domain $-\infty < x \leq 0$ comprises a two region structure, namely an outer region for $-x=O(1)$ and an inner region for $-x=O(\mu)$.

\subsection{The outer solution}

For $-x=O(1)$ we pose the regular expansions\footnote{It may be verified later from the leading order inner solution that there are no terms intruding in this expansion between the leading and first order terms.}
\begin{eqnarray}
    v(x) &=& v_0(x) + \mu^4 v_1(x) + o(\mu^4), \hspace{0.5cm} \nonumber \\
    \rho_r(x) &=& \rho_{r0}(x) + \mu^4 \rho_{r1}(x) + o(\mu^4), \hspace{0.5cm} \nonumber \\ 
    \rho_{m}(x) &=& \rho_{m0}(x) + \mu^4 \rho_{m1}(x) + o(\mu^4), \hspace{1cm} \mbox{as $\mu \to 0$},  
\label{eq:outerseries}
\end{eqnarray}
to give the leading order problem 
\be
  \rho_{r0} + \rho_{m0} = \frac{3 x^2}{\kappa^2} , \hspace{0.5cm} \frac{d \rho_{r0}}{dx} = \frac{4v_0 \rho_{r0}}{x} , 
\hspace{0.5cm} \frac{d \rho_{m0}}{dx} = \frac{3v_0 \rho_{m0}}{x}, 
\label{eq:outge0}
\ee
subject to the conditions 
\be
   \mbox{as $x \to -\infty$} \hspace{1cm} v_0 \sim 1/2, \hspace{0.25cm} \rho_{r0} \sim \frac{3x^2}{\kappa^2} . 
\label{eq:outbc0}
\ee
At first order $O(\mu^4)$ we obtain  
\begin{eqnarray}
 && -6 \sigma x\kappa^2(\rho_{r1} + \rho_{m1})(2v_0-1)^3 = \left( \frac{d v_0}{dx} - \frac{3}{2x}(1-2v_0)(2-v_0)v_0 \right)  , \label{eq:outge1a} \\
 &&  \frac{d \rho_{r1}}{dx} = \frac{4}{x}(v_0 \rho_{r1} + v_1 \rho_{r0} ), 
\hspace{0.5cm} \frac{d \rho_{m1}}{dx} = \frac{3}{x}(v_0 \rho_{m1} + v_1 \rho_{m0} ) . 
\label{eq:outge1b}
\end{eqnarray}
with 
\be
   \mbox{as $x \to -\infty$} \hspace{1cm} v_1 =o(1), \hspace{0.25cm} \rho_{r1} = o(1) . 
\label{eq:outbc1}
\ee

\subsubsection{The leading order solution}

Eliminating $\rho_{m0}$ in (\ref{eq:outge0}) gives  
\be
     v_0 = \frac{6x^2}{9x^2 + \kappa^2 \rho_{r0}} ,  
\label{eq:v0}
\ee 
with $\rho_{r0}$ satisfying  the nonlinear first-order ODE 
\be
     x \frac{d \rho_{r0}}{dx} = \frac{ 24 x^2  \rho_{r0} }{ (9x^2 + \kappa^2 \rho_{r0})} .  
\ee
This last equation is scale invariant under the transformation $x=\alpha\bar{x}, \rho_{r0}=\alpha^2 \bar{\rho_{r0}}$ for real $\alpha$, suggesting the change of variables $x=-e^{y},\rho_{r0}=x^2u(y)$ which gives the autonomous equation
\[
      \frac{du}{dy} = \frac{2u(3-\kappa^2 u)}{(9+\kappa^2 u)} ,  
\]
with general solution 
\[
     3 \left| \frac{\kappa^2 u}{k_0} \right |^{3/4} = (3-\kappa^2 u) e^{y/2} ,  
\]
where $k_0>0$ is an arbitrary constant of integration (and the multiplicative constants being introduced for later convenience). 
In terms of the original variables this gives the implicit solution 
\be
       3 x^2 - \kappa ^2 \rho_{r0}  = 3 \left( \frac{\kappa^2 \rho_{r0}}{k_0 } \right)^{3/4} , 
\label{eq:rhor0}
\ee
with $v_0$ then being given by (\ref{eq:v0}) and $\rho_{m0}$ being determined from the first equation in (\ref{eq:outge0}). For reference we note the following limiting behaviours of the leading order outer solution, namely 
\be
\hspace{-2.5cm} \mbox{as $x \to -\infty$ } \hspace{1cm}  v_0 &\sim& \frac{1}{2} + \frac{1}{8} \left(\frac{3}{k_0}\right)^{3/4} (-x)^{-1/2} - \frac{1}{16}\left(\frac{3}{k_0}\right)^{3/2} (-x)^{-1}, \nonumber
\\
\hspace{-1cm} \rho_{r0} &\sim& \frac{3}{\kappa^2} x^2  - \frac{3}{\kappa^2} \left(\frac{3}{k_0}\right)^{3/4} (-x)^{3/2} + \frac{9}{4\kappa^2} \left(\frac{3}{k_0}\right)^{3/2} (-x),  \nonumber 
\\
\hspace{-1cm} \rho_{m0} &\sim&  \frac{3}{\kappa^2} \left(\frac{3}{k_0}\right)^{3/4} (-x)^{3/2} - \frac{9}{4\kappa^2} \left(\frac{3}{k_0}\right)^{3/2} (-x),  
\label{eq:outff}
\ee
whilst 
\be
     \mbox{as $x \to 0^-$ } \hspace{1cm} v_0 \sim \frac{2}{3} \left( 1 - \frac{k_0}{9} (-x)^{2/3} \right) , 
      \hspace{0.25cm} \rho_{r0} \sim \frac{k_0}{\kappa^2} (-x)^{8/3} ,  
    \hspace{0.25cm} \rho_{m0} \sim  \frac{3}{\kappa^2}x^2 - \frac{k_0}{\kappa^2} (-x)^{8/3} . 
\label{eq:outx0}
\ee
As expected, the leading order outer solution has no dependence on $\sigma$.

We note that the outer problem has no inherent time scale, so that the scalings  
\[
    x = k_0^{-3/2} \bar{x} , \hspace{0.5cm} v = \bar{v}, \hspace{0.5cm} \rho_{r} = k_0^{-3} \bar{\rho}_{r}, 
\hspace{0.5cm} \rho_{m} =k_0^{-3} \bar{\rho}_{m},
\]
remove the constant $k_0$, so that without loss of generality we may consider $k_0=1$.

\begin{figure}[t!]
\begin{center}
\epsfig{figure=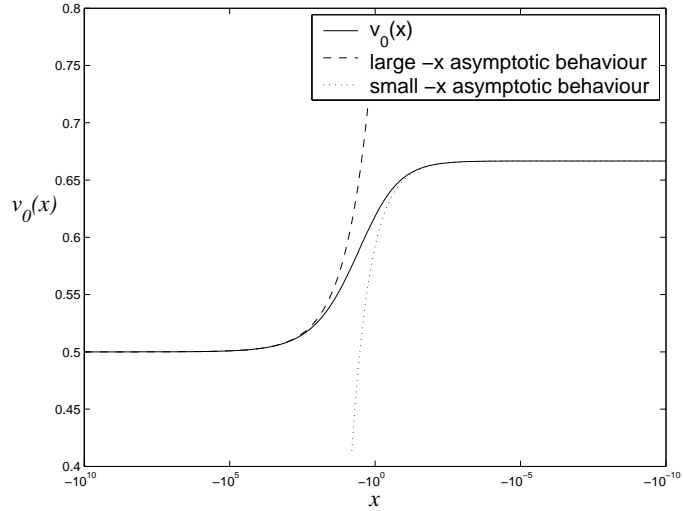, width=9cm}
\end{center}
\caption{ \small    Plot of the leading order outer solution $v_0(x)$ using (\ref{eq:v0}) and (\ref{eq:rhor0}) 
over a range of $x<0$ values in the parameter case $\kappa^2=k_0=1$. Also shown are the large and small $-x$ asymptotic behaviours for $v_0$ in (\ref{eq:outff}) and (\ref{eq:outx0}).  
 }
\label{figouter}
\end{figure}

\subsubsection{The first order solution}

Using the leading order solution for $v_0$ in (\ref{eq:outge1a}) gives the simplification 
\be
          \rho_{m1} = -\frac{\sigma v_0 (3 v_0 + 2) }{12 \kappa^2 x^2 (2v_0-1)^2} - \rho_{r1} , 
\label{eq:rhom1}
\ee
and then eliminating $\rho_{m1}$ in (\ref{eq:outge1b}) gives 
\be
     v_1 = \frac{\sigma v_0^3(39 v_0-2)}{72 x^4(2v_0-1)^2} - \frac{v_0^2}{6x^2} \kappa^2 \rho_{r1} . 
\label{eq:v1}
\ee
Consequently the equation for $\rho_{r1}$ in (\ref{eq:outge1b}) then gives 
\be
    x \frac{d \rho_{r1}}{dx} = 6 v_0^2 \rho_{r1} + \frac{\sigma v_0^2(2-3v_0)(39v_0-2)}{6\kappa^2x^2(2v_0-1)^2} .
\label{eq:derhor1}
\ee
The use of an integrating factor allows us to obtain a quadrature expression for $\rho_{r1}$, however we simply note that 
this is possible rather than record it since its form will not be explicitly used. Using the behaviour (\ref{eq:outff}) for $v_0$ we may deduce from 
(\ref{eq:derhor1}) the behaviour 
\be
    \kappa^2 \rho_{r1} \sim  \frac{7 \sigma}{3} \left( \frac{k_0}{3} \right)^{3/2} x^{-1} 
    - \frac{17\sigma}{12} \left( \frac{k_0}{3} \right)^{3/4} (-x)^{-3/2} 
    +   \sigma k_1 (-x)^{3/2} \left( 1 - \frac{3}{2} \left( \frac{3}{k_0} \right)^{3/4} (-x)^{-1/2} + O( x^{-1})\right) 
\\ \mbox{as $x \to -\infty$} \nonumber , 
\ee
where $k_1$ is an arbitrary constant. The requirement (\ref{eq:outbc1}) enforces $k_1=0$, thus setting the homogeneous solution terms to zero. Consequently from 
(\ref{eq:rhom1}) and (\ref{eq:v1}) we have that 
\be
\kappa^2 \rho_{m1} \sim -\frac{7\sigma}{4} \left( \frac{k_0}{3} \right)^{3/4} (-x)^{-3/2}, \hspace{0.25cm} 
v_{1} \sim \frac{7\sigma}{12} \left( \frac{k_0}{3} \right)^{3/2} (-x)^{-3}    \hspace{1cm}
\mbox{as $x \to -\infty$} .
\ee
Similarly using (\ref{eq:outx0}) in (\ref{eq:derhor1}) with (\ref{eq:rhom1}) and (\ref{eq:v1}), we have that  
\be
  \mbox{as $x \to 0^-$} \hspace{1cm}  \kappa^2 \rho_{r1} \sim -\frac{8\sigma k_0}{9} (-x)^{-4/3}, 
    \hspace{0.25cm}  \kappa^2 \rho_{m1} \sim  - 2 \sigma x^{-2} , 
 \hspace{0.25cm}  v_{1} \sim \frac{8\sigma }{9} x^{-4} .  
\label{eq:outfirstx0}
\ee
 
\subsection{The inner solution}

In the outer region, the derivative of $v$ in Eqn.~(\ref{eq:fullv}) is neglected at leading order. This is not expected to hold uniformly as $x \to 0^-$, since dominant balance in 
Eqn.~(\ref{eq:fullv}) is given when $-x=O(\mu)$ (when using the scalings $v=O(1),\rho_r=O(\mu^{8/3}),\rho_m=O(\mu^{2})$ from the limiting behaviour (\ref{eq:outx0})). Consequently we now consider an inner (or boundary layer) region at $-x=O(\mu)$, for which we introduce the scaled variables
\be
    x= \mu X , \hspace{0.25cm} v = V , \hspace{0.25cm} \rho_{r} = \mu^{8/3} P_r, \hspace{0.25cm} \rho_{m} = \mu^{2} P_m . 
\label{eq:innervars}
\ee   
The system (\ref{eq:fullv})--(\ref{eq:fullmat}) then becomes for $-X=O(1)$ 
\begin{eqnarray}
  && \frac{dV}{dX} = \frac{3(1-2V)(2-V)V}{2X} - 6\sigma X (2V-1)^3 \left( \kappa^2(P_m + \mu^{2/3}P_r) -3X^2\right) , \\
  && \frac{d P_{m}}{dX} = \frac{3 V P_{m}}{X}, \\
  &&  \frac{d P_{r}}{dX} = \frac{4 V  P_{r}}{X} . 
\end{eqnarray}
For the inner expansion, we pose 
\be
    V(X)= V_0(X) + O(\mu^{2/3}), \hspace{0.25cm} P_m(X)= P_{m0}(X) + O(\mu^{2/3}), \hspace{0.25cm} P_{r}(X) = P_{r0}(X) + O(\mu^{2/3}), 
\hspace{1cm} \mbox{as $\mu \to 0$},  
\label{eq:innerseries}
\ee
to obtain the leading order equations 
\begin{eqnarray}
  && \frac{dV_0}{dX} = \frac{3(1-2V_0)(2-V_0)V_0}{2X} - 6\sigma X (2V_0-1)^3 \left( \kappa^2 P_{m0} -3X^2\right) , \label{eq:inge1} \\
  && \frac{d P_{m0}}{dX} = \frac{3 V_0 P_{m0}}{X}, \label{eq:inge2}\\
  &&  \frac{d P_{r0}}{dX} = \frac{4 V_0  P_{r0}}{X} , \label{eq:inge3}
\end{eqnarray}
subject to the leading order outer matching conditions
\be
    \mbox{as $X \to -\infty$} \hspace{1cm} V_0 \sim \frac{2}{3} , \hspace{0.25cm} P_{m0} \sim \frac{3}{\kappa^2}(-X)^{2} , 
     \hspace{0.25cm} P_{r0} \sim \frac{k_0}{\kappa^2}(-X)^{8/3} , 
\label{eq:innerff}
\ee
which follow from (\ref{eq:outx0}). It is noteworthy that $P_{r0}$ does not enter (\ref{eq:inge1}), illustrating that it does not affect the leading order behaviour of $V(X)$ in this inner region, only its correction terms. Again $k_0=1$ may be considered without loss of generality since the scalings   
\[
    X =\bar{X} , \hspace{0.5cm} V_0 = \bar{V}_0, \hspace{0.5cm} P_{r0} = k_0 \bar{P}_{r0}, 
\hspace{0.5cm} P_{m} = \bar{P}_{m0},
\]
remove the constant $k_0$ from the leading order inner problem.   

\begin{figure}[t!]
\begin{center}
\epsfig{figure=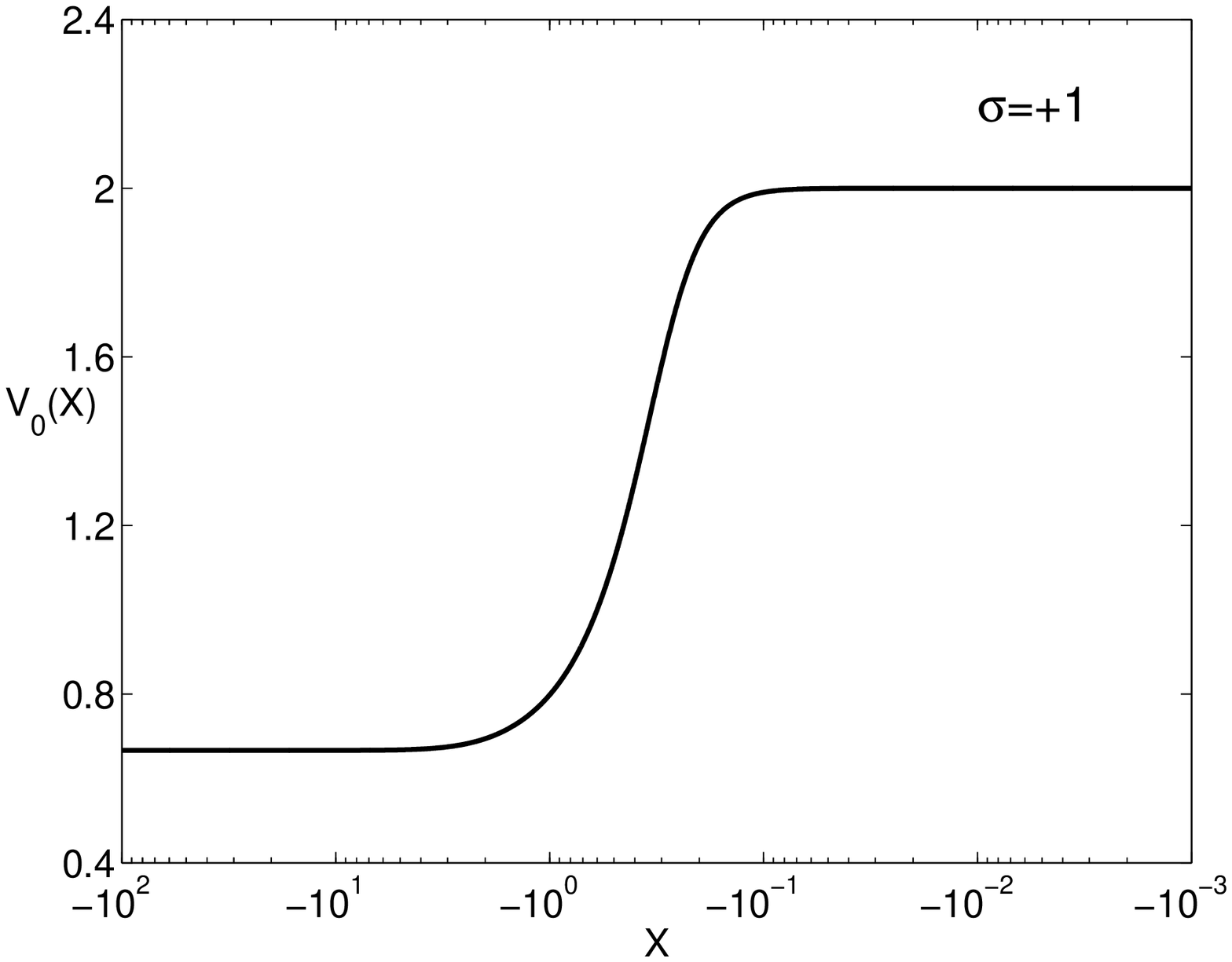, width=7cm}
\epsfig{figure=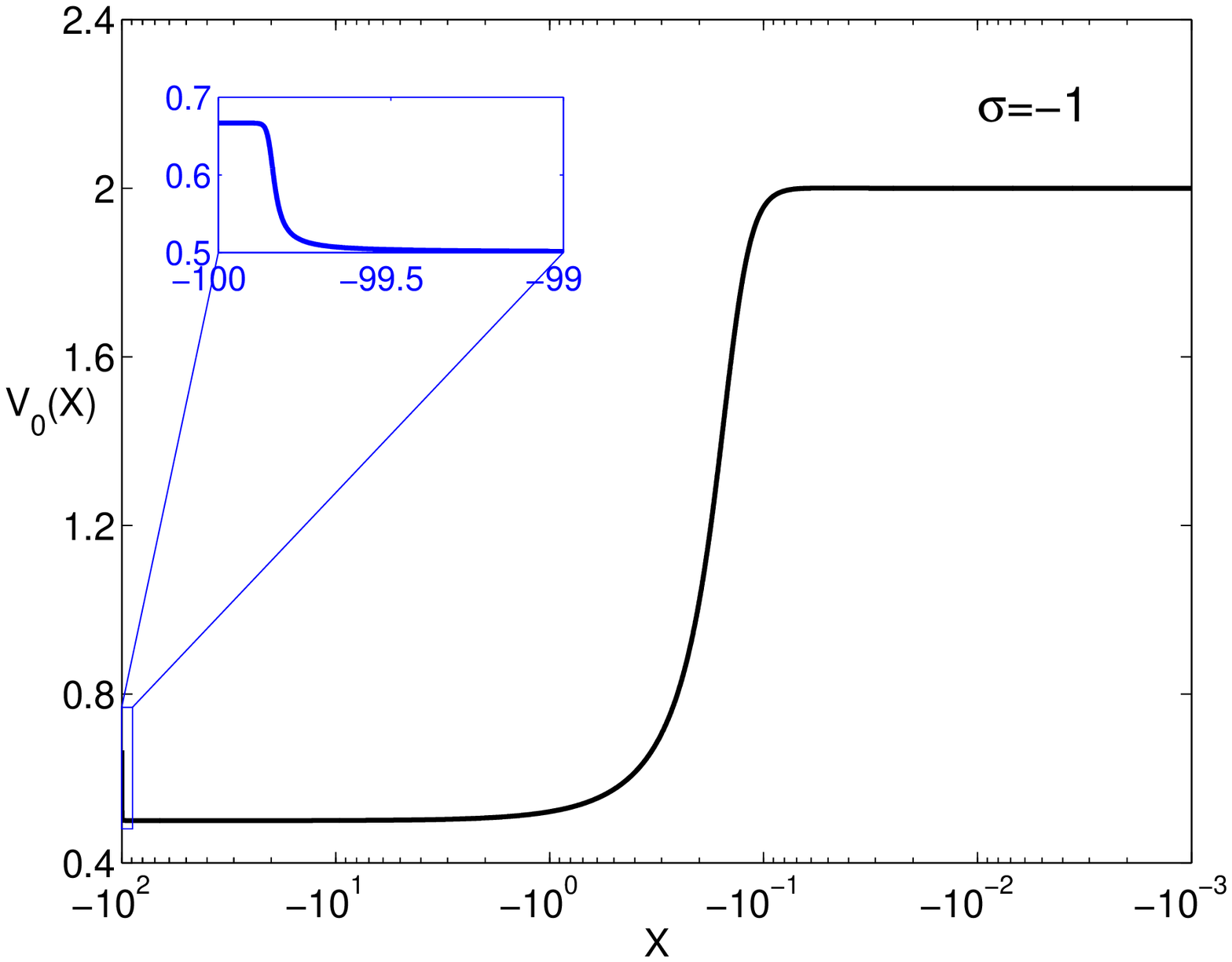, width=7cm}
\end{center}
\caption{ \small
Illustration of the numerical solution to the leading order inner problem (\ref{eq:inge1})--(\ref{eq:inge3}) subject to (\ref{eq:innerff})
as an initial-value-problem solved over $X \in [X_{inf},X_0]$ with $X_{inf}=-10^{2},X_0=-10^{-3}$ in the parameter case $\kappa^2=k_0=1$. Both CDTT ($\sigma=1$) and mCDTT ($\sigma=-1$) models are considered.}
    
\label{figinner}
\end{figure}

An eigenmode analysis given in Appendix \ref{app:emodes} about the far-field asymptotic behaviour (\ref{eq:innerff}) for the system (\ref{eq:inge1})--(\ref{eq:inge3}), confirms that 
(\ref{eq:innerff}) impose the necessary three conditions on the nonlinear third order system. Consequently this leading order inner problem may be solved as an initial-value-problem (IVP). Using (\ref{eq:inge1})--(\ref{eq:inge3}), a more accurate 
far-field expansion consistent with (\ref{eq:innerff}) is 
\be
    \mbox{as $X \to -\infty$} \hspace{1cm} V_0 \sim \frac{2}{3} + \frac{8\sigma}{9}X^{-4}, \hspace{0.25cm} P_{m0} \sim \frac{3}{\kappa^2}(-X)^{2} e^{-\frac{2\sigma}{3X^4}} , 
     \hspace{0.25cm} P_{r0} \sim \frac{k_0}{\kappa^2}(-X)^{8/3} e^{-\frac{8\sigma}{9X^4}} . 
\label{eq:innerff2}
\ee
Using these asymptotic expansions, Eqn (\ref{eq:inge1}) becomes
\be
\frac{dV_0}{dX} \sim \frac{40}{9}\frac{\sigma}{X^5}
\label{eq:dvdxsim}
\ee
and we see that the slope of the asymptotic behaviour depends on $\sigma$.
Numerical solution of (\ref{eq:inge1})--(\ref{eq:inge3}) subject to (\ref{eq:innerff2}) was obtained using MATLAB's IVP solver ode15s, using solver tolerances of AbsTol=RelTol$=10^{-13}$. The 
system was solved over the interval $X \in [X_{inf},X_0]$ where $X_{inf}=-10^2,X_0=-10^{-3}$, the condition (\ref{eq:innerff2}) being imposed at $X=X_{inf}$. The parameter values $\kappa^2=k_0=1$ where used. Figure \ref{figinner} illustrates the behaviour of $V_0$, for the cases $\sigma=\pm 1$.

A local analysis near $V_0=2$ for the system (\ref{eq:inge1})--(\ref{eq:inge3}) gives the behaviours 
\be
    \mbox{as $X \to 0^-$} \hspace{1cm} V_0 \sim 2 - \frac{486\sigma}{5}X^{4}, \hspace{0.25cm} P_{m0} \sim A_{m0} X^{6} e^{-\frac{729 \sigma X^4}{10}} , 
     \hspace{0.25cm} P_{r0} \sim A_{r0} X^{8} e^{-\frac{486 \sigma X^4}{5}} , 
\label{eq:innerX0}
\ee
for constants $A_{m0},A_{r0}$ (which can be determined numerically $A_{m0}\approx 98, A_{r0}\approx 105$ in the case $\kappa^2=k_0=1$ and $\sigma=1$). 

\begin{figure}[t!]
\begin{center}
\epsfig{figure=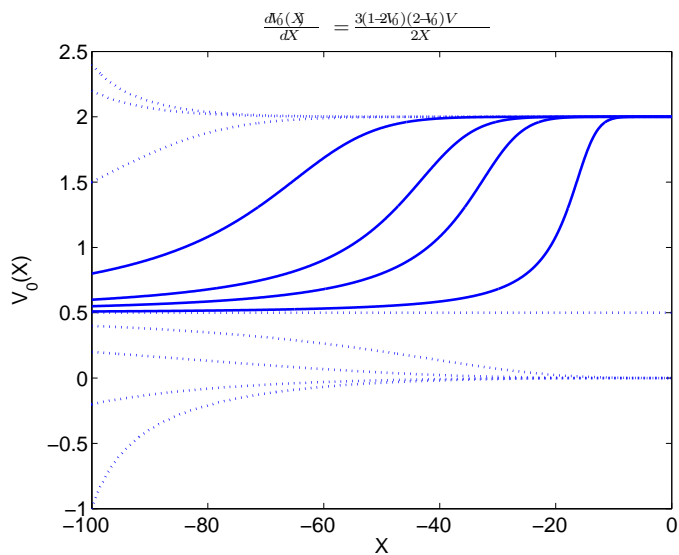, width=9cm}
\end{center}
\caption{ \small
The phase plane of possible solution branches of the leading order inner problem, Eqn.~(\ref{eq:V0small}), which is independent of $\sigma$.  We have included negative values of $v$ for completeness, although we never consider such values cosmologically.  Notice that the solutions are symmetric around $X\rightarrow-X$.}
\label{V0phase}
\end{figure}

For $-X \ll 1$, (\ref{eq:inge1}) simplifies to  
\be
\frac{dV_0}{dX} = \frac{3(1-2V_0)(2-V_0)V_0}{2X} ,
\label{eq:V0small}
\ee
a phase plane of possible solution branches of which being given in Figure \ref{V0phase}. 
The numerical solution of Figure \ref{figinner} enters the stable equilibrium point $V_0=2$ along one of the branches shown in
$\left\{-\infty<X<0,1/2<V_0 \leq 2 \right\}$.

The dependence of (\ref{eq:dvdxsim}) on $\sigma$ is crucial for the inner solution.  
For the CDTT model, the positive slope ensures that the solution moves from $V_0\sim 2/3$ ($X\to -\infty$) to the final solution $V_0\sim2$ ($X\to 0^-$).  
However, when $\sigma=-1$, the asymptotic behaviour around $V_0\sim 2/3$ has negative slope and the solution rapidly approaches the solution, $V_0=1/2$.  
We believe this is the ``recent radiation epoch" behaviour found in \cite{Amendola:2006kh,Fairbairn:2007qp}, which is not cosmologically viable.  
As $X\to 0^-$, the numerical solution then follows that of Figure \ref{V0phase} and $V_0\to 2$.  
In fact, the $\sigma=-1$ (CDTT) model cannot accomodate the outer solution of $v_0\sim2/3$, but rather must match $v_0=1/2$ for the leading order outer solution.
As
such, then, we do not obtain a viable asymptotic solution in the $\sigma=-1$ case.
In the remaining sections of this paper, we only consider the mCDTT model, which has a cosmologically viable solution.

\subsection{Matching and composite expansion of the mCDTT model}

For completeness we now illustrate matching between the two term outer solution and the one term inner solution for the mCDTT model, before considering a composite asymptotic expansion valid for all $x\leq 0$.

Taking two terms of the outer expansion (\ref{eq:outerseries}) and using the behaviours (\ref{eq:outx0}) and (\ref{eq:outfirstx0}), we have for small $-x$ that 
\begin{eqnarray}
    &&  v(x) = v_0(x) + \mu^4 v_1(x) + o(\mu^4) \sim \frac{2}{3}  - \frac{2k_0}{27} (-x)^{2/3} + \mu^4 \frac{8}{9}x^{-4}   ,  \nonumber  \\  
  &&  \rho_r(x) = \rho_{r0}(x) + \mu^4 \rho_{r1}(x) + o(\mu^4) \sim  \frac{k_0}{\kappa^2} (-x)^{8/3} - \mu^4 \frac{8k_0}{9\kappa^2}(-x)^{-4/3} ,   \nonumber \\  
  &&  \rho_m(x) = \rho_{m0}(x) + \mu^4 \rho_{m1}(x) + o(\mu^4) \sim \frac{3}{\kappa^2}x^2 - \frac{k_0}{\kappa^2} (-x)^{8/3} - \mu^4 \frac{2}{\kappa^2}x^{-2} ,    \nonumber
\end{eqnarray}  
which in terms of inner variables (\ref{eq:innervars}) becomes
\begin{eqnarray}
    &&  V(X) \sim \frac{2}{3} + \frac{8}{9} X^{-4} + O(\mu^{2/3})  ,  \nonumber \\  
  &&  P_r(X) \sim \frac{k_0}{\kappa^2} (-X)^{8/3} - \frac{8k_0}{9 \kappa^2} (-X)^{-4/3} ,  \nonumber  \\  
  &&  P_m(X) \sim \frac{3}{\kappa^2}X^2 -  \frac{2}{\kappa^2} X^{-2} + O(\mu^{2/3}),  \label{eq:outin}  
\end{eqnarray}  
Taking the inner expansion (\ref{eq:innerseries}) and expanding (\ref{eq:innerff2}) for large $-X$ gives 
\begin{eqnarray}
  && V(X) = V_0(X) + O(\mu^{2/3}) = \frac{2}{3} + \frac{8}{9}X^{-4} + o(X^{-4}) + O(\mu^{2/3}) , \nonumber \\ 
  && P_{r}(X) = P_{r0}(X) + O(\mu^{2/3}) = \frac{k_0}{\kappa^2}(-X)^{8/3} \left(1 -\frac{8}{9X^4} \right) + o((-X)^{-4/3}) + O(\mu^{2/3}) , \nonumber \\ 
 && P_{m}(X) = P_{m0}(X) + O(\mu^{2/3}) =  \frac{3}{\kappa^2} X^{2} \left(1 -\frac{2}{3X^4} \right) + o(X^{-2}) + O(\mu^{2/3}) , \label{eq:inff}
\end{eqnarray}
these expressions agreeing with those in (\ref{eq:outin}) to the order of the terms retained (i.e. the first two terms). 

\begin{figure}[t!]
\begin{center}
\epsfig{figure=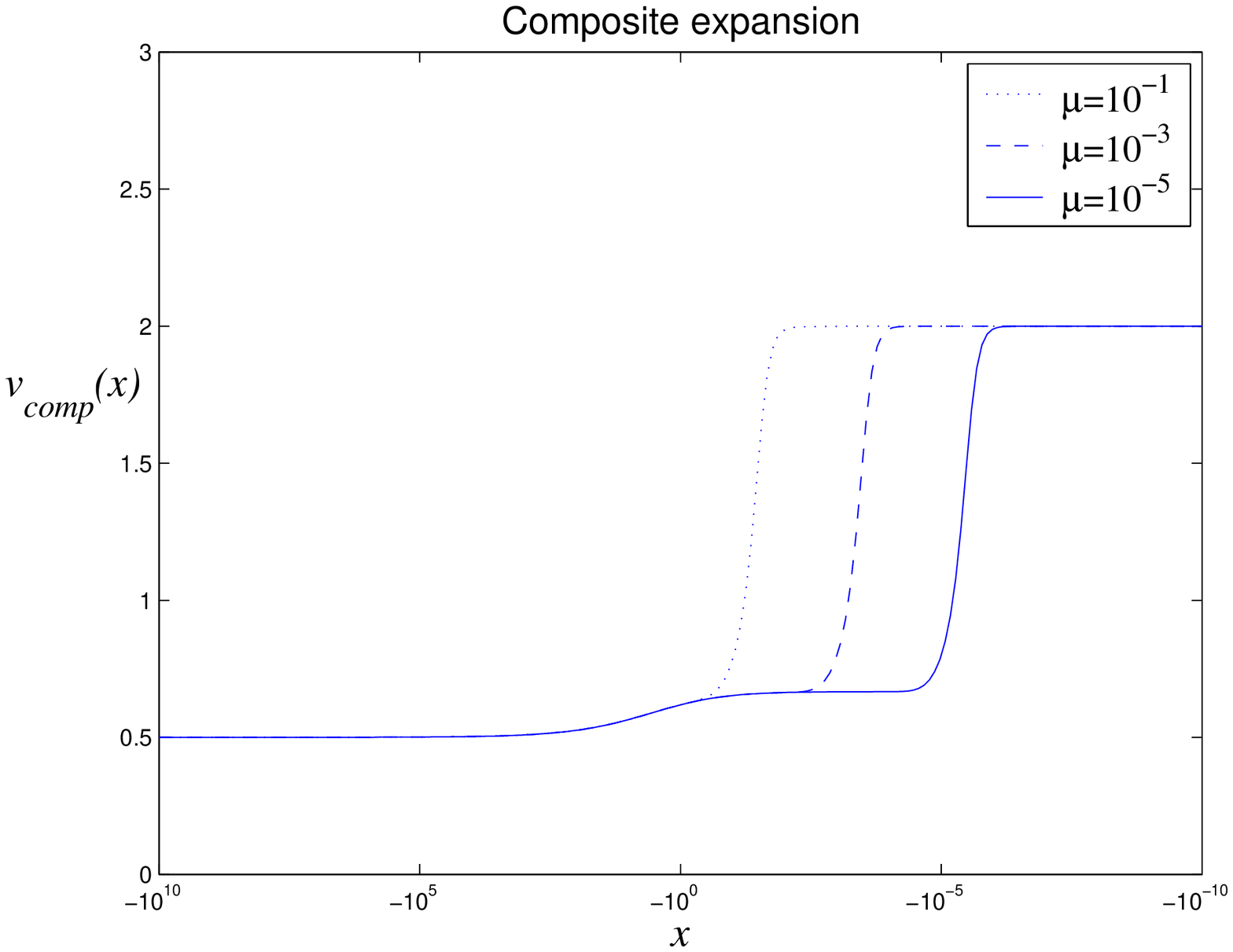, width=9cm}
\end{center}
\caption{ \small
Illustration of the leading order composite expansion when $\sigma=1$ (mCDTT model)} for $v$ given in (\ref{eq:vcomp}) for selected $\mu$ values.    
\label{figvcomp}
\end{figure}

A composite expansion using the leading order terms of the outer and inner expansions is given by 
\begin{eqnarray}
   && v_{\rm comp}(x) =  v_0(x) + V_0 \left( \frac{x}{\mu} \right) - \frac{2}{3}, \hspace{0.25cm} \label{eq:vcomp} \\
  && \rho_{r {\rm comp}}(x) =  \rho_{r0}(x) + P_{r0} \left( \frac{x}{\mu} \right) - \frac{k_0}{\kappa^2} (-x)^{8/3}, \hspace{0.25cm} \\ 
  && \rho_{m {\rm comp}}(x) =  \rho_{m0}(x) + P_{m0} \left( \frac{x}{\mu} \right) - \frac{3}{\kappa^2} x^{2}, \label{eq:rhomcomp}
\end{eqnarray}
where the last terms in each expression are the leading order terms in the overlap region between $-x=O(1)$ and $-x=O(\mu)$. Such expressions are now uniformly valid throughout the interval $-\infty <x \leq 0$.  A plot of $v_{\rm comp}(x)$ is given in Figure \ref{figvcomp} for three selected values $\mu=10^{-1},10^{-3},10^{-5}$ illustrating the emergence of a two step profile as $\mu$ decreases.

\section{Cosmological Implications}
\label{sec:relevance}

Since there is no observation that can probe the future expansion of the universe, we are restricted to testing the validity of the outer solution, which describes the past:
\begin{eqnarray}
    v(x)&=& v_0(x) + \mu^4 v_1(x) + o(\mu^4).
\end{eqnarray}
This can be done in several ways, but we restrict ourselves to just two:  the requirement that we have a long enough period of matter domination to satisfy large scale structure and that we satisfy the constraints of Big Bang Nucleosynthesis (BBN).  We restrict ourselves to the mCDTT model, which we have seen leads to a standard cosmic history.

We define an ``initial" epoch, given by $x_0$, at which the ratio of radiation to matter is set:
\begin{eqnarray}
r_0 = \frac{\rho_{r}(x_0)}{\rho_{m}(x_0)}.
\end{eqnarray}
We can relate the parameter $k_0$ to this ratio,
\begin{eqnarray}
k_0 = 3\left(r_0+1\right)^{4/3}(-x_0)^{-\frac23}
\label{eq:koratio}
\end{eqnarray}
Setting $k_0=\kappa^2=1$ (as in Section \ref{sec:asymp}), we find matter-radiation equality ($r_0=1$) at $x_0 = -12\sqrt{3}\approx -20.78$, as can be seen in Figure \ref{figvcomp}.

In comparison to data, we must convert the ``timescale'' from $H$ ($x$) to redshift, $z$:
\begin{eqnarray}
1+z = \frac{a_0}{a(t)} 
\end{eqnarray}
where $a_0$ is the scale factor at present times.  We define this to be the transition time from $v=\frac23$ (such that the universe is just accelerating today) and acts as a normalisation.  We note that 
\be
d\ln a = H dt = - \frac{v}{x} dx.
\label{eq:dadx}
\ee

\subsection{Sufficient Matter Domination}
The redshift of equality, $z_{\rm eq}$, is found from the first peak in the Cosmic Microwave Background (CMB) anisotropy spectrum.  From the WMAP first year data~\cite{Spergel:2003cb}, $z_{\rm eq} = 3454^{+385}_{-392}$ (WMAP data only).  In order to ensure a long enough epoch of matter domination, we require that $v(x)\approx 2/3$ up to this redshift.

We may relate the redshift of matter domination back to the free parameter $k_0$.  
We may use Eqn.~(\ref{eq:dadx}) and require that, at the present time, $x(0)=-H_0 = -\mu$ and  $\ln a_0=0$.  $H_0$ is the present value of the Hubble parameter.  Assuming that $v=\frac23$, we find the relation:
\be
x = x_{0} \left(\frac{a_{0}}{a}\right)^{3/2} = -\mu \left(1+ z\right)^{3/2}.
\label{eq:xz}
\ee
At radiation-domination equality, $x_{\rm eq} \approx -2 \times 10^5 \mu$.   
Using Eqn.~(\ref{eq:koratio}) and defining $k_0$ at $x_{\rm eq}$, we find that, to ensure a long enough period of matter domination:
\be
k_0 \approx 0.002 \mu^{-2/3}.
\label{eq:k0}
\ee

Despite appearances, we note that this solution agrees with the analysis of~\cite{Amendola:2006we}, although we quantify the period of matter domination and find a sufficiently long era.  
We note that, during this matter epoch, due to the asymptotic solution, $\Omega_m \to 1$, but never reaches (nor sits at) unity.
In Amendola et al's notation, the solution undergoes a transition from the critical point, $P_5$ to $P_6$, where $m_5 \to 0^-$, $m_6=-2$ and $m^{\prime}_{5,6}>-1$.  
This transition is allowed according to~\cite{Amendola:2006we}.
At the transition, such variables ($m$ etc.) become infinite, due to the factor of $1/(1+\tilde f)$.  

\begin{figure}[t!]
\begin{center}
\epsfig{figure=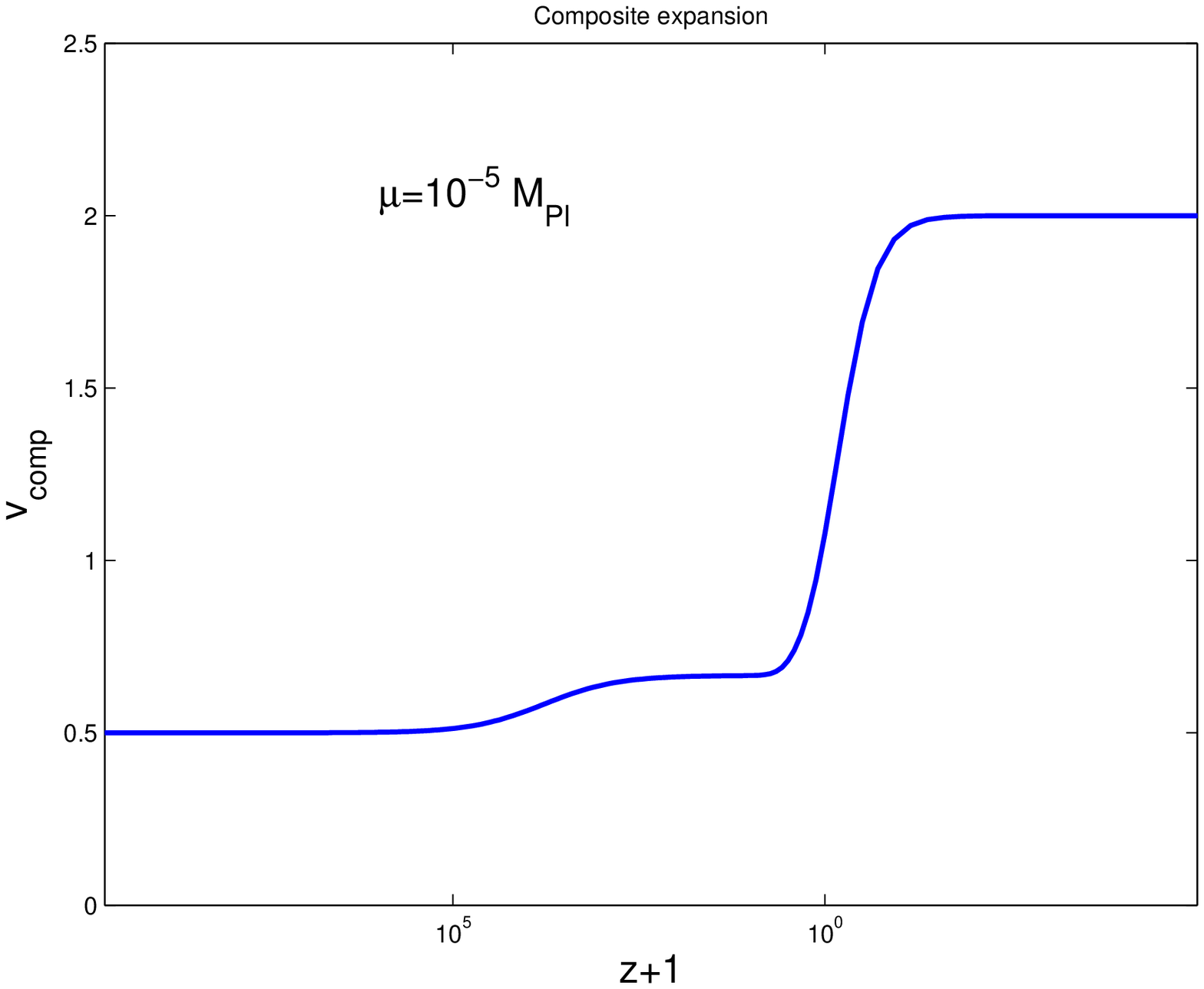, width=9cm}
\end{center}
\caption{ \small
The composite solution, $v_{\rm comp}$ shown against redshift, $z$, in the mCDTT model ($\sigma=1$)} when $\mu=10^{-5} \mpl$. In order to obtain sufficient matter domination, we require $k_{0}=4.3\mpl^{-2/3}$.   
\label{figcomp_red}
\end{figure}

\subsection{Big Bang Nucleosynthesis}
One of the most constraining tests of non-standard cosmological expansion rates comes from the earliest epoch we can test: 
Big Bang Nucleosynthesis.  In an early radiation-dominated epoch, the expansion rate is determined by the energy density of the relativistic particles.  Due to this, tight experimental constraints exist.  (For a review, see \cite{Kneller:2004jz}).  Firstly, we note that at the time of BBN, $z_{\rm BBN}=5.9\times 10^{9}$, and using Eqn.~(\ref{eq:xz}),
\be
\frac{x_{\rm BBN}}{x_{0}} = -\frac{x_{\rm BBN}}{\mu} = \left(5.9\times10^{9}\right)^{3/2}.
\ee

One parameterisation of the BBN constraint is given as 
\be
\rho_{r} \rightarrow \rho_r^{\prime} = \rho_{r} + \rho_{X} = S^{2}\rho_{r} \nonumber
\ee
where $\rho_{X}$ denotes the extra component, in our context due to gravity.  The constraint parameter, $S$, can be written in terms of the addition of an extra neutrino species,
\be
S^{2}=1+7/43\Delta N_{\nu} < 1.16,
\ee
where, in the last term, we have assumed that $\Delta N_{\nu}\leq1$~\cite{Fields:1996yw}.  In the mCDTT model, with the asymptotic solution found in Section \ref{sec:asymp}, we find
\be
S^{2}-1&=&\frac{\mu^{4}\rho_{r1}}{\rho_{r0}} \nonumber \\
     &\approx& \mu^{4}\left[-\frac{8}{9} \frac{k_{0}}{\kappa^{2}} \left(-x_{\rm BBN}\right)^{-4/3}\right] \left[\frac{3x_{\rm BBN}^{2}}{\kappa^{2}}+ ...\right]^{-1} \nonumber \\
     &\approx& -\frac{8}{27}\mu^{4}k_{0}\left(-x_{\rm BBN}\right)^{-10/3}  \nonumber
\ee
Substituting the value of $k_{0}$ required for sufficient matter domination, Eqn.~(\ref{eq:k0}), we find
\be
 S^{2}-1  \approx - 5.9\times 10^{-4} \left(\frac{\mu}{-x_{\rm BBN}}\right)^{10/3} \approx -8.25 \times 10^{-53}, \nonumber
\ee
which is obviously well within present constraints.

\section{Generalisation}
\label{sec:gen}
It must be acknowledged that the mCDTT model may ruled out by observation within the solar system and therefore is perhaps not the best model to consider.  We note, however, that it provides a useful example of the technique and illustrates at least one cosmologically viable $f(R)$ model.  We now wish to consider other models for which this technique is valid, so to identify a class of models which might work fully.

We firstly note that for any model with $R+f(R)$, the system can be termed as in Eqn.~(\ref{eq:generaldiff}) and therefore the leading order solutions, given by Eqns.~(\ref{eq:rhor0}-\ref{eq:outx0}), are always true.  This is to be expected, since we want to pick up the Einstein solution when $\mu\to 0$.  Due to this solution, we also expect to have the same inner scaling for $v$, $\rho_r$ and $\rho_m$, as given in Eqn.~(\ref{eq:innervars}):
\be
 v = V , \hspace{0.25cm} \rho_{r} = \epsilon^{2/3} P_r, \hspace{0.25cm} \rho_{m} = \epsilon^{1/2} P_m . \nonumber
\ee
where we have substituted $\epsilon=\mu^4$ to account for a more general parameterisation.
With these scalings, the inner problem for a general $f(R)$ theory can be written:
\be
&& \frac{dV}{dx} = -\frac{V^2(V-1)}{6 x^3}\frac{\tilde f_R}{\epsilon^{1/2}\tilde f_{RR}} 
     - \frac{V^3}{36 x^5}\frac{\tilde f}{\epsilon\tilde f_{RR}} + \frac{2V(2V-1)}{x}
     + \frac{\left[3x^2 -\kappa^2\left(P_m +\mu^{2/3}P_r\right)\right]V^3}{18x^5~\epsilon^{3/2}\tilde f_{RR}} \label{eq:gendV}\\
   && \frac{d P_{m}}{dX} = \frac{3 V P_{m}}{X}, \nonumber \\
   && \frac{d P_{r}}{dX} = \frac{4 V  P_{r}}{X} . \nonumber 
\ee
The relevant scaling is suggested from $\epsilon^{1/2}\tilde f=O(1)$ (although this is not the only condition).
For inverse power-law models, 
\be
\tilde f(R)=AR^{-n}
\ee
with positive $n$, we find the scaling  $x= \epsilon^{1/4n} X$ so that
\be
\frac{dV}{dX} &=& -\frac{V(V-1)(2V-1)}{(n+1)X}
     - \frac{V(2V-1)^2}{n(n+1)X}
     + \frac{2V(2V-1)}{X}
     + \frac{2(6^n)V^{1-n}(2V-1)^{n+2} X^{2n-1}\left[3X^2 -\kappa^2\left(P_m +\mu^{2/3}P_r\right)\right]}{n(n+1)A} \nonumber \\
 &=& \frac{V(2V-1)(n+2)}{n(n+1)X}\left[V-\frac{(n+1)(2n+1)}{(n+2)}\right]  
     + \frac{2(6^n)V^{1-n}(2V-1)^{n+2}X^{2n-1}\left[3X^2 -\kappa^2\left(P_m +\mu^{2/3}P_r\right)\right]}{n(n+1)A} \nonumber 
\ee
We thereby note that as $X\to -\infty$, the leading order term is the last term (proportional to $X^{2n-1}$) and we expect to recover the Einstein solution.  As $X\to0^-$, the solution is dominated by the first terms.  We expect dominant balance when $-x=O(\epsilon^{1/4n})$ and therefore sufficient matter domination requires
\be
k_{0}=0.002\epsilon^{-1/6n}.
\ee
The (accelerating) asymptotic solution as $X\to 0^-$, which we denote by $V_{\rm att}$,  can be calculated from
\be
V_{\rm att} = \frac{(n+1)(2n+1)}{(n+2)}.
\label{eq:Vatt}
\ee
For $n=1$, we find $V_{\rm att}=2$ as seen in Figure~\ref{figvcomp}.
We strongly note, however, that the sign of the coefficient, $A$, seems crucial if one wishes to avoid the apparent $\phi$MDE solution of \cite{Amendola:2006kh}.  In each model, a full matching analysis is required.

\section{Conclusion}
\label{sec:conc}

A modification of gravity has been suggested in the form of $f(R)$ theories.  However, until now, it has been believed that these cannot produce a standard era of matter domination.
By studying the asymptotic behaviour of a class of $f(R)$ models, we show that a sufficiently-long period of matter domination can be achieved.

We consider one specific example, the (m)CDTT model, $R+\sigma\mu^4/R$ with $\sigma=\pm1$, and explicitly calculate the asymptotic behaviour.  
In terms of the CDDETT variables, where $x=-H$, we find that the system gives a singular perturbation problem, comprising a two region structure: an outer region $-x=O(1)$ and an inner region $x=O(\mu)$.
We show that the mCDTT solution recovers Einstein cosmology when $-x=O(1)$ (i.e. in the past when $H\gg\mu$) and matches an accelerated cosmology when $-x=O(\mu)$ (i.e. today when $H\sim\mu\sim H_0$).
We directly relate this solution to cosmological variables and find the condition for which we obtain sufficient matter domination to satisfy large scale structure data.  We also show that the perturbation causes no significant effect at the time of Big Bang Nucleosynthesis.

We generalise our results to include the class of inverse power-law functions.  Although these are also ruled out by solar system tests, we aim to exhibit the generality of our previous results, which could be valid for other functions.
We find that the original CDTT model cannot match to the outer $v\sim2/3$ (matter dominated) solution and the ``recent radiation'' behaviour of \cite{Amendola:2006kh,Fairbairn:2007qp} is found.
Generally, we find that for models with $R\pm \epsilon AR^{-n}$, standard cosmological evolution is expected until $-x=O(\epsilon^{1/4n})$, which can be coincident with the present time.  After this point, an accelerated solution is found.  This solution depends strongly on the sign of $A$.

Our main result is that, for a model with lagrangian $R+f(R)$, we expect the leading order outer (past-time) solution to be given by Eqn.~{\ref{eq:rhor0}}, the Einstein solution.  
The next order terms will be proportional to the perturbative parameter in $f(R)$.
The leading order inner (future) solution can be found using the system in Eqn.~(\ref{eq:gendV}).  We therefore expect the solutions to match across the boundary and the universe should find the accelerating attractor, given by Eqn.~(\ref{eq:Vatt}).  Whether the ``recent radiation epoch'' is found (instead of a matter epoch) seems to depend on the sign of the corrective term.

It is the authors' belief that this perturbation method can be applied to even more general functions of higher curvature invariants, for example $f(R,P,Q)$, where $P=R_{\mu\nu}R^{\mu\nu}$ and $Q=R_{\alpha\beta\gamma\delta}R^{\alpha\beta\gamma\delta}$, such as the CDDETT model.  This is the focus of future work.  With solutions such as we have shown here, observational constraints can be applied to a much larger class of modified gravity models than is possible at present.

\appendix

\section{Eigenmode analysis (mCDTT model)}
 \label{app:emodes}

The number of degrees of freedom contained within the asymptotic behaviour (\ref{eq:innerff}) may be determined through an eigenmode analysis. 
We consideration a perturbation about the far-field behaviour in the form 
\be
   V_0(X) \sim \frac{2}{3} + \delta \hat{V}(X), \hspace{0.5cm}   P_{m0}(X) \sim \frac{3}{\kappa^2} X^2 + \delta \hat{P}_{m}(X), \hspace{0.5cm}   
  P_{r0}(X) \sim \frac{k_0}{\kappa^2} (-X)^{8/3} + \delta \hat{P}_{r}(X),  \hspace{1cm} \mbox{as $X \to -\infty$} .
\ee
Keeping terms at $O(\delta)$, (\ref{eq:inge1})--(\ref{eq:inge3}) give the linearised equations 
\begin{eqnarray}
  && X \frac{d \hat{V}}{dX} + 3 \hat{V} + \frac{2\sigma}{9} X^2 \kappa^2 \hat{P}_m = 0  , \label{eq:lin1} \\
  && X \frac{d \hat{P}_{m}}{dX} - 2  \hat{P}_{m} - \frac{9X^2}{\kappa^2}\hat{V} = 0 , \label{eq:lin2}\\
  && X \frac{d \hat{P}_{r}}{dX} -  \frac{8}{3} \hat{P}_{r} - \frac{4 k_0 (-X)^{8/3}}{\kappa^2}\hat{V} = 0 , \label{eq:lin3}
\end{eqnarray}
Eliminating $\hat{P}_m$ from (\ref{eq:lin1}) and (\ref{eq:lin2}) gives the transformed Bessel equation
\[
     X^2\frac{d^2 \hat{V}}{d X^2} +  4X(1-\sigma)\frac{d \hat{V}}{dX} + 2(X^4-6) \hat{V} = 0
\]
with solutions 
\[
     \hat{V} =  \left\{ \begin{array}{l} 
                        (-X)^{m} J_{\nu}(X^2/\sqrt{2}) ,  \\
                      (-X)^{m} Y_{\nu}(X^2/\sqrt{2}) ,
                      \end{array}
                \right. 
\]
where 
\begin{eqnarray}
m=2\sigma-3/2, \quad \nu = \frac14 \sqrt{57-24\sigma+16\sigma^2}
\end{eqnarray}
where $J_{\nu}(z),Y_{\nu}(z)$ are standard first kind Bessel functions of order $\nu$.  Their large $-X$ behaviours are  
\[
     \hat{V} \sim   \left\{ \begin{array}{l} 
                        (-X)^{l}\cos \left( \frac{X^2}{\sqrt{2}} - {\cal S}\pi\right) ,  \\
                      (-X)^{l} \sin \left( \frac{X^2}{\sqrt{2}} - {\cal S}\pi\right) ,
                      \end{array}
                \right.  \hspace{1cm} \mbox{as $X \to -\infty$},
\] 
where
\begin{eqnarray}
l=2\sigma-5/2, \quad {\cal S} = \frac{\left(2\nu+1\right)}{4}
\end{eqnarray}
(to within multiplicative constants) and consequently the three linearly independent eigenmodes are 
\be
    \left. \begin{array}{l}
             \hat{V} \sim (-X)^{l}\cos \left( \frac{X^2}{\sqrt{2}} - {\cal S}\pi\right)  \\
             \hat{P}_m \sim  \frac{9}{\sqrt{2} \kappa^2} (-X)^{l}\sin \left( \frac{X^2}{\sqrt{2}} - {\cal S}\pi\right)\\ 
             \hat{P}_r \sim \sim \frac{2\sqrt{2} k_0}{ \kappa^2} (-X)^{l+2/3}\sin \left( \frac{X^2}{\sqrt{2}} - {\cal S}\pi\right)
 \end{array} \right\} ; \hspace{0.25cm} 
    \left. \begin{array}{l}
             \hat{V} \sim -(-X)^{l}\sin \left( \frac{X^2}{\sqrt{2}} - {\cal S}\pi\right)  \\
             \hat{P}_m \sim  \frac{9}{\sqrt{2} \kappa^2} (-X)^{l}\cos \left( \frac{X^2}{\sqrt{2}} - {\cal S}\pi\right)\\ 
             \hat{P}_r \sim \sim \frac{2\sqrt{2} k_0}{ \kappa^2} (-X)^{l+2/3}\cos \left( \frac{X^2}{\sqrt{2}} - {\cal S}\pi\right)
 \end{array} \right\} ; \hspace{0.25cm} 
    \left. \begin{array}{l}
             \hat{V} = 0   \\
             \hat{P}_m = 0  \\ 
             \hat{P}_r = (-X)^{8/3} 
 \end{array} \right\} .  
\label{eq:emodes}
\ee
The third mode in (\ref{eq:emodes}) corresponds to small changes in $k_0$ and is the only mode consistent with the 
far-field behaviour (\ref{eq:innerff}). This implies that there is only one degree of freedom in the 
asymptotic behaviour (\ref{eq:innerff}) which is associated with $k_0$. Thus if $k_0$ is specified then the far-field behaviour 
imposes three conditions on the system (\ref{eq:inge1})--(\ref{eq:inge3}).

\section*{Acknowledgements}
LH wishes to thank Anthony Brookfield, Carsten van de Bruck and Damien Easson for many useful discussions.  LH also thanks Elizabeth Winstanley for comments that lead to this project. LH also wishes to acknowledge and thank STFC for funding.

\bibliographystyle{h-physrev3}
\bibliography{paper}

\end{document}